%% file: contribution.tex
{  

\documentclass[a4paper,11pt]{article}

\usepackage{contribution}
\usepackage{epsfig}

\input{econfmacros}
\input{contributionmacros}

\begin{document}

\contribution[]  
{A new perspective on the $\Delta_{5/2^{+}}(2000)$ puzzle}  
{P. }{Gonz\'alez}  
{$^1$Departamento de F\'{\i}sica Te\'{o}rica (Universidad de
Valencia (UV)), Valencia, Spain \\
$^2$Instituto de F\'\i sica Corpuscular (IFIC), Centro Mixto
CSIC-Universidad de Valencia, Institutos de Investigaci\'on de
Paterna, Aptd. 22085, E-46071 Valencia, Spain \\
$^3$Department of Physics, Zhengzhou University, Zhengzhou, Henan 450001, China\\
$^4$Yukawa Institute for Theoretical Physics, Kyoto University, Kyoto 606-8502, Japan}  
{}  
{$^{1,2}$, Ju-Jun Xie$^{2,3}$, A. Mart\'inez Torres$^4$, and E.Oset$^{1,2}$}  

%

\abstract{%
We argue that $\Delta_{5/2^{+}}(2000)(\ast\ast),$ cataloged as a
resonance in the Particle Data Book Review (PDG), should be
interpreted instead as two
distinctive resonances, $\Delta_{5/2^{+}}(\sim1740)$ and $\Delta_{5/2^{+}%
}(\sim2200).$ Our argument is based on a solution of the
$\pi\Delta\rho$ problem in a Fixed Center Approximation (FCA) to the
Fadeev equations. $\Delta_{5/2^{+}}(\sim1740)$ can then be
interpreted as a $\pi-(\Delta \rho)_{N(1675)}$ bound state. As a
corollary $\Delta_{1/2^{+}}(1750)(\ast)$ can be understood as a $\pi
N_{1/2^{-}}(1650)$ bound state.
}
%

\section{Introduction}

In the PDG \cite{pdg2010} there is only a well established
$\Delta_{5/2^{+}}$ resonance, $\Delta(1905)$ $F_{35}$
($\ast\ast\ast\ast),$ and fair evidence of the existence of another
one, $\Delta(2000)$ $F_{35}$ ($\ast\ast)$. However, a careful look
at this last resonance shows that its nominal mass is in fact
estimated from $\Delta(1724\pm61)$, $\Delta(1752\pm32)$ and $\Delta
(2200\pm125)$ respectively extracted from three independent analyses
\cite{Vra00,Man92,Cut80}. Moreover a recent new data analysis has
reported a $\Delta_{5/2^{+}}$ with a pole position at $1738$ MeV
\cite{Suz10}.

From a $3q$ description the $\Delta_{5/2^{+}}(1905)$ is naturally
accommodated as the lowest $\Delta_{5/2^{+}}$ state in the second
energy band \cite{capstick}. Similarly, the reported
$\Delta_{5/2^{+}}(2200)$ with a more uncertain mass ($2200\pm125$
MeV) may be reasonably located in the fourth energy band. On the
contrary $\Delta_{5/2^{+}}(1740)$ can not be accommodated as a $3q$
state without seriously spoiling the overall spectral description.
The same kind of problem was tackled in \cite{Pej09} regarding the
description of $\Delta_{5/2^{-}}(1930)$ with a mass much lower than
the corresponding to the third energy band, the first available band
for such a state. There the consideration of the $\rho\Delta$
channel, whose threshold ($2002$ MeV) lies close above the
experimental mass of the resonance and far below the $3q$ mass
($\sim2150$ MeV), allowed for an explanation of
$\Delta_{5/2^{-}}(1930)$ and its partners, $\Delta_{3/2^{-}}(1940)$
and $\Delta_{1/2^{-}}(1900),$ as
$\rho\Delta$ bound states in the $I=3/2$ sector. In addition $N_{1/2^{-}%
}(1650),$ $N_{3/2^{-}}(1700)$ and $N_{5/2^{-}}(1675)$ were also well
described as $\rho\Delta$ bound states in the $I=1/2$ sector
although with a bigger sensitivity in this case to the cutoff (or
subtraction constant) parameter employed in the chiral unitary
approach.

For $\Delta_{5/2^{+}}(1740)$ one can easily identify a meson-baryon
threshold, $\left[  \pi N_{5/2^{-}}(1675)\right]  _{threshold}=1814$
MeV, in between the $3q$ mass ($\sim1910$ MeV) and data. Then one
can wonder about the possibility that the $\pi N_{5/2^{-}}(1675)$
system may give rise to a bound state which could provide
theoretical support to the fair evidence of the existence of
$\Delta_{5/2^{+}}(1740)$. Actually this bound state nature would
provide an explanation to the extraction of this resonance in some
data analyses and not in others. It turns out that only analyses
reproducing the $\pi\pi N$ production cross section data extract it.
Let us note that this would be a necessary condition to extract
$\Delta_{5/2^{+}}(1740)$ if corresponding to a $\pi
N_{5/2^{-}}(1675)$ state (let us recall that $N_{5/2^{-}}(1675)$
decays to $\pi N$ and to $\pi\pi N$ with branching fractions of 40\%
and 55\% respectively). To examine this possibility we have
performed an analysis of the $\pi N_{5/2^{-}}(1675)$ system by
assuming that $N_{5/2^{-}}(1675)$ is a $\rho\Delta$ bound state (we
have fixed the subtraction constant $a_{\Delta\rho}=-2.28$ to get
precisely the mass of $N(1675)$) \cite{Xie11}.

\section{Formalism}

The interaction of a particle with a bound state of a pair of
particles at very low energies or below threshold can be efficiently
and accurately studied by means of the FCA to the Faddeev equations
for the three-particle system \cite{Chan62}. For the
$\Delta$-$\rho$-$\pi$ system the $\pi$ is assumed to scatter one by
one the fixed centers $\rho$ and $\Delta$. Then the total three-body
scattering amplitude $T$ is given in terms of two partition
functions $T_{1}$ and $T_{2}$ accounting for the diagrams starting
with the interaction of the $\pi$ with particle $i$ of the compound
system:
\begin{align}
T & = T_{1}+T_{2}, \\
T_{1} &  =t_{1}+t_{1}G_{0}T_{2},\\
T_{2} &  =t_{2}+t_{2}G_{0}T_{1}%
\end{align}
where $t_{i}$ represent the $\Delta\pi$ and $\rho\pi$ unitarized
scattering amplitudes, see Refs.~\cite{Sar05,Roc05} for details.
$G_{0}$ is
the loop function for the $\pi$ meson propagating inside the $N_{5/2^{-}%
}(1675)$ resonance (see Ref.~\cite{Xie11} for details).

\section{Results and discussion}

The dynamic generation of resonances from our formalism depends on
two subtraction constants$,$ $a_{\Delta\pi}$ and $a_{\rho\pi},$
respectively associated to the $\Delta$-$\pi$ and $\rho$-$\pi$
unitarized $s-$wave interactions entering in our calculation. We
assume that they are effective parameters, their values implicitly
taking into account the effects of the $3q$ component of
$N_{5/2^{-}}(1675).$ Therefore they could differ significantly from
the values used in Refs.~\cite{Sar05,Roc05}.

\begin{figure}[htbp]
\begin{center} \vspace{-0.3cm}
\includegraphics[scale=0.4]{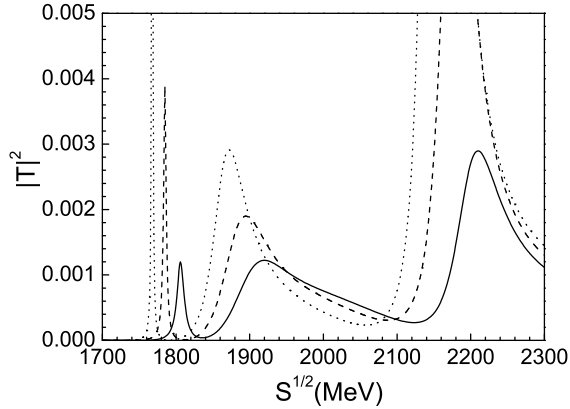} \vspace{-0.2cm}
\caption{Modulus squared of the three-body scattering amplitude for
$I=3/2$. Results obtained with $a_{\rho\pi}=-2.0$ and
$a_{\Delta\pi}=-2.6$ (solid line), $-3.0$ (dash line), $-3.4$
(dotted line). \label{fig:3half}}
\end{center}
\end{figure}

Examples of results giving rise to a $\pi-N(1675)$ bound state are
graphically shown in Fig. \ref{fig:3half}. The three peaks in the
figures may be unambiguously assigned to $\Delta _{5/2^{+}}(1740),$
$\Delta_{5/2^{+}}(1905)$ and $\Delta_{5/2^{+}}(2200).$ Note that the
location of the first peak varies from $1770$ MeV to $1800$ MeV, a
little bit higher than the masses of the existing candidates in \cite{pdg2010}%
. Indeed we could force $a_{\Delta\pi}=-4.3$ to get an average mass
of $1740$ MeV. Therefore these results should mainly be interpreted
as a fit to fix the parameters in our formalism. In order to gain
confidence about the possible existence of $\Delta_{5/2^{+}}(1740)$
it becomes essential that further predictions from our formalism
(with no free parameters) are successful in the interpretation of
data. Let us examine some of these predictions in the $I=3/2$
sector.

$\Delta$ resonances generated from $\pi N_{3/2^{-}}(1700)$ and $\pi
N_{1/2^{-}}(1650)$ are of particular interest since
$N_{3/2^{-}}(1700)$ and $N_{1/2^{-}}(1650)$ are dynamically
generated from $\Delta\rho$ as degenerate
states to $N_{5/2^{-}}(1675).$ Hence we predict $I=3/2,$ $J^{P}=1/2^{+}%
,3/2^{+}$ states almost degenerate to $\Delta_{5/2^{+}}(1740).$ For
$J^{P}=1/2^{+}$ the state may be assigned to
$\Delta_{1/2^{+}}(1750)(\ast),$ a resonance that can not be
described by quark models \cite{capstick}. It should be remarked
that only analyses reproducing the $\pi\pi N$ production cross
section data extract it as it was the case for
$\Delta_{5/2^{+}}(1740)$. Therefore the mere existence of
$\Delta_{1/2^{+}}(1750)(\ast)$ could be considered within our
calculation framework as an argument in favor of the
existence of $\Delta_{5/2^{+}}(1740).$ In what respects to $\Delta_{3/2^{+}%
}(\sim1740)$ it is assigned to $\Delta_{3/2^{+}}(1600)$ which is
also overpredicted by $3q$ models as the first radial excitation of
the $\Delta(1232).$ However, other channels as $\sigma\Delta(1232)$
and $\pi N_{3/2^{-}}(1520)$ could be playing a more important role
in the generation of this resonance  \cite{Gon07}.

Additional analyses have been done in the $I=3/2,1/2$ sectors
\cite{Xie11}. The consistency of the whole scheme and the good
agreement with data makes us confident that the approximations
followed draw the essential dynamics. From our results we may
conclude that there is a sound theoretical basis to support
the data analyses extracting two distinctive resonances, $\Delta_{5/2^{+}%
}(1740)$ and $\Delta_{5/2^{+}}(2200),$ cataloged altogether as
$\Delta _{5/2^{+}}(2000)$ in the PDG.

%

\acknowledgements{%
This work is partly supported by DGICYT Contract No. FIS2006-03438,
the Generalitat Valenciana in the project PROMETEO and the EU
Integrated Infrastructure Initiative Hadron Physics Project under
contract RII3-CT-2004-506078. J.-J.X. acknowledges Ministerio de
Educaci\'{o}n Grant SAB2009-0116. The work of A.M.T. is supported by
the Grant-in-Aid for the Global COE Program \textquotedblleft The
Next Generation of Physics, Spun from Universality and Emergence"
from the Ministry of Education, Culture, Sports, Science and
Technology (MEXT) of Japan. P.G. benefits from the funding by the
Spanish Ministerio de Ciencia y Tecnolog\'{\i}a and UE FEDER under
Contract No. FPA2007-65748, by the Spanish Consolider Ingenio 2010
Program CPAN (CSD2007-00042) and by the Prometeo Program (2009/129)
of the Generalitat Valenciana. Partial funding is also provided by
HadronPhisics2, a FP7-Integrating Activities and Infrastructure
Program of the EU under Grant 227431. }


%

\end{document}

%% file: econfmacros.tex


\newcommand{\weblink}[2][]{%
    \ifthenelse{\equal{#1}{}}%
    {\textnormal{\url{#2}}}%
    {\textnormal{\href{#2}{#1}}}%
}

\newcommand{\acknowledgements}[1]{%
  \bigskip\bigskip
  \textsf{\textbf{\Large Acknowledgements}} \\[2ex]
  {#1}
  \bigskip
}


\def\beq{\begin{equation}}
\def\eeq#1{\label{#1}\end{equation}}
\def\eeqn{\end{equation}}

\def\beqa{\begin{eqnarray}}
\def\eeqa#1{\label{#1}\end{eqnarray}}
\def\eeqan{\end{eqnarray}}



\let\bar=\overbar





\def\Dslash{\not{\hbox{\kern-4pt $D$}}}
\def\dslash{\not{\hbox{\kern-2pt $\del$}}}


\def\msb{{\bar{\ssstyle M \kern -1pt S}}}


%

%% file: contributionmacros.tex

\newcommand{\contribution}[7][]{%
  \clearpage
  \thispagestyle{plain}
  \ifthenelse{\equal{#1}{}}
  {\hypersetup{pdftitle={#2}}}
  {\hypersetup{pdftitle={#1}}}
  \hypersetup{pdfauthor={{#3} {#4}}}
  {\centering\normalfont\LARGE\bfseries\sffamily #2 \par\nobreak}
  \lhead{}
  \chead{%
    \textit{\footnotesize XIV International Conference on Hadron Spectroscopy
      (\weblink[\textit{hadron2011}]{http://www.hadron2011.de}), 13-17 June 2011, Munich, Germany}%
  }
  \rhead{}
  \bigskip
  \begin{center}
    {#3} {#4}\ifthenelse{\equal{#6}{}}{}{\footnote{\weblink[#6]{mailto:#6}}}
    \ifthenelse{\equal{#7}{}}{}{#7} \\
    \textit{#5}
  \end{center}
  \bigskip
}

\renewcommand{\abstract}[1]{%
  \begin{center}
    \begin{minipage}{0.85\textwidth}
      \begin{footnotesize}
        #1
      \end{footnotesize}
    \end{minipage}
  \end{center}
  \bigskip
}

%